\documentclass[aps, 11pt, tightenlines, floatfix, onecolumn, amsmath, superscriptaddress, showkeys]{revtex4}
\usepackage{graphicx}
\usepackage{dcolumn}
\usepackage{bm}
\usepackage{float}
\usepackage{hyperref}

\begin{document}

\title{Estimating $\gamma\gamma$ absorption for UHE photons with lepton and hadron production}

\author{Giorgio Galanti}
\email{gam.galanti@gmail.com}
\affiliation{INAF, Osservatorio Astronomico di Brera, Via E. Bianchi 46, I -- 23807 Merate, Italy}

\author{Fulvio Piccinini}
\email{fulvio.piccinini@pv.infn.it}
\affiliation{INFN, Sezione di Pavia, Via A. Bassi 6, I -- 27100 Pavia, Italy}

\author{Fabrizio Tavecchio}
\email{fabrizio.tavecchio@inaf.it}
\affiliation{INAF, Osservatorio Astronomico di Brera, Via E. Bianchi 46, I -- 23807 Merate, Italy}

\author{Marco Roncadelli}
\email{marco.roncadelli@pv.infn.it}
\affiliation{INFN, Sezione di Pavia, Via A. Bassi 6, I -- 27100 Pavia, Italy}
\affiliation{INAF, IASF Milano, Via A. Corti 12, I -- 20133 Milano, Italy}

\date{\today}
\begin{abstract}
Surprisingly, the contribution to the cosmic opacity to UHE ($E>10^{18} \, \rm eV$) $\gamma$-rays has been systematically computed so far for the $\gamma \gamma \to e^+e^-$ and $\gamma \gamma \to e^+ e^- e^+ e^-$
processes alone. We go a step further by systematically evaluating the additional opacity brought about by other leptons and hadrons. We find that the dominant channels are those leading to the production of $\mu^{\pm}$ and hadrons (mainly $\pi^{\pm}$, $\pi^0$, $K^{\pm}$, $K^0$, $\eta$). For nearly the GZK radius, the photon survival probability becomes smaller by about a factor of two with respect to current estimates.
\end{abstract}



\maketitle


\section{Introduction}

The propagation of $\gamma$-rays in the Universe is severely limited by absorption through the interaction with soft radiation backgrounds, leading to the production of  particle/antiparticle
pairs \cite{tau1}. The most commonly considered reaction is the $e^{\pm}$ pair production 
$\gamma\gamma\to e^+e^-$, which starts to become important above few tens of GeV for sources
at cosmological distances. Below energies $\simeq \, 10^{14} \, {\rm eV}$ the main targets are the IR-optical-UV photons belonging to the so called extragalactic background light (EBL), the fossil record of light produced by
stars along the whole cosmic history (for a review, see~\cite{Dwek}). At energies above $\simeq 10^{14} \, {\rm eV}$ the major source of opacity is instead the cosmic microwave background (CMB), which restricts 
the mean free path of $\gamma$-rays of energy $\simeq 10^{15}$ eV to few tens of Megaparsecs
(see e.g.~\cite{dgrTransp}). Ultra-high-energy (UHE) $\gamma$-rays, with energies above $10^{18} \, {\rm eV}$, interact with the Rayleigh-Jeans tail of the CMB spectrum and the cosmic radio background (RB). The potential detection of UHE photons is quite relevant, since they provide a natural probe for several fundamental processes~\cite{Gelmini1,auger}. In particular, UHE photons are the by-product of the photo-meson reactions which restrict the propagation of UHE cosmic rays to $\simeq 100$ Mpc (the so-called GZK radius)~\cite{Gelmini2}. Currently, only upper bounds on the UHE flux have been derived~\cite{auger}. Because photons produced at cosmic distances suffer from absorption during their propagation to the Earth, any comparison of these upper limits with theoretical expectations must rely upon the accurate estimate
of the cosmic opacity at UHE \cite{Gelmini2,TauDPP}.

So far, the optical depth $\tau_{\gamma}$ at UHE has been modeled by using specific codes~\cite{TauDPP} which, however, include only pure QED effects (and often consider only the $\gamma\gamma \to e^+e^-$ process). The aim of this Letter is to quantify the relevance of hadron-producing and lepton-producing reactions, which are possible for photons in such an energy range because of the large energies available
in the center of mass (CM) frame ${\cal S}_{\rm CM}$. These processes have been considered rather cursorily  many years ago within the framework of effective treatments (see e.g.~\cite{originalDDP} and references therein). Here, we compute them in a systematic fashion and in the light of the state-of-the-art knowledge.


\section{Pair production (PP) cross section}

\begin{figure}      
\begin{center}
\includegraphics[width=.4\textwidth]{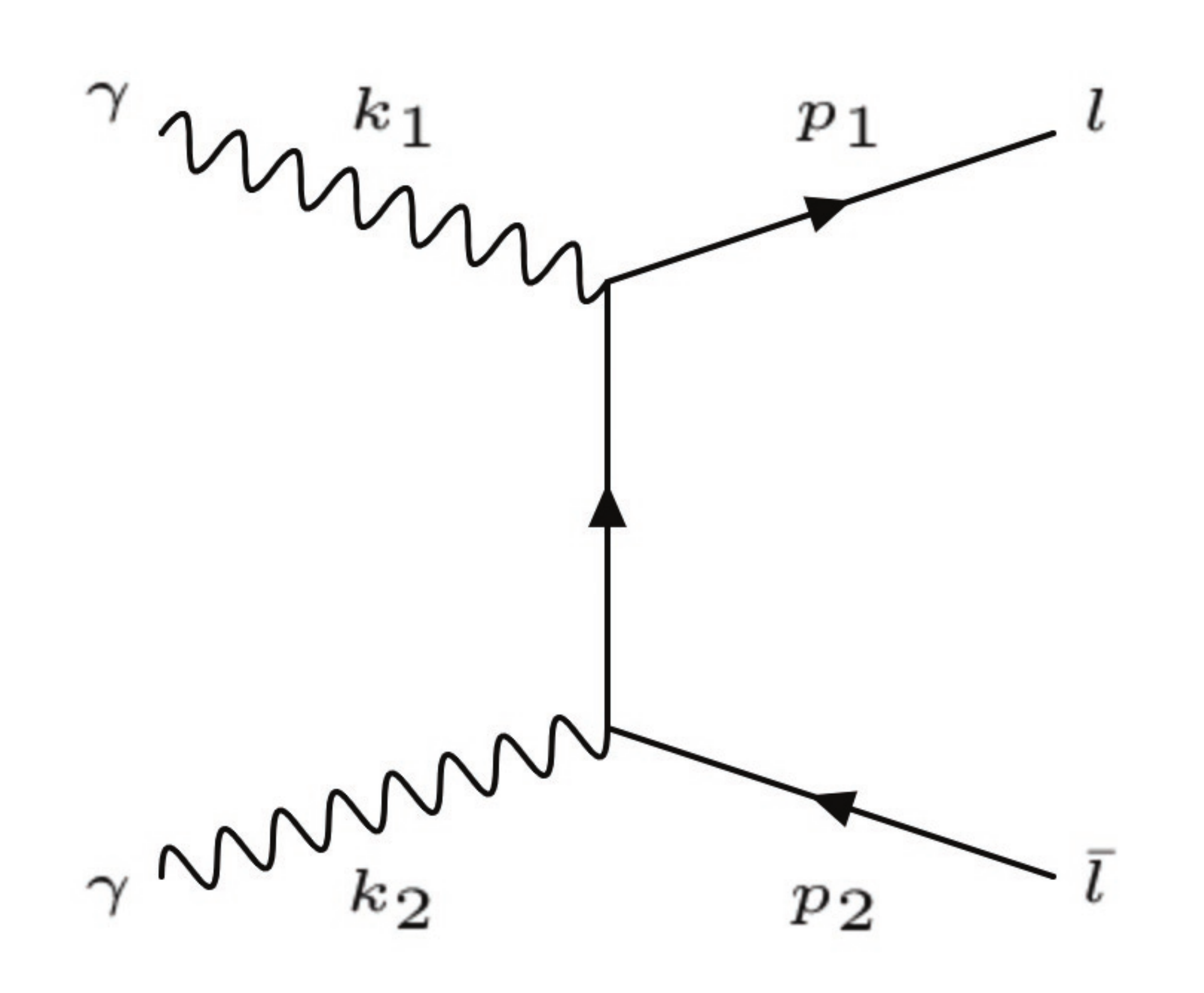}\includegraphics[width=.4\textwidth]{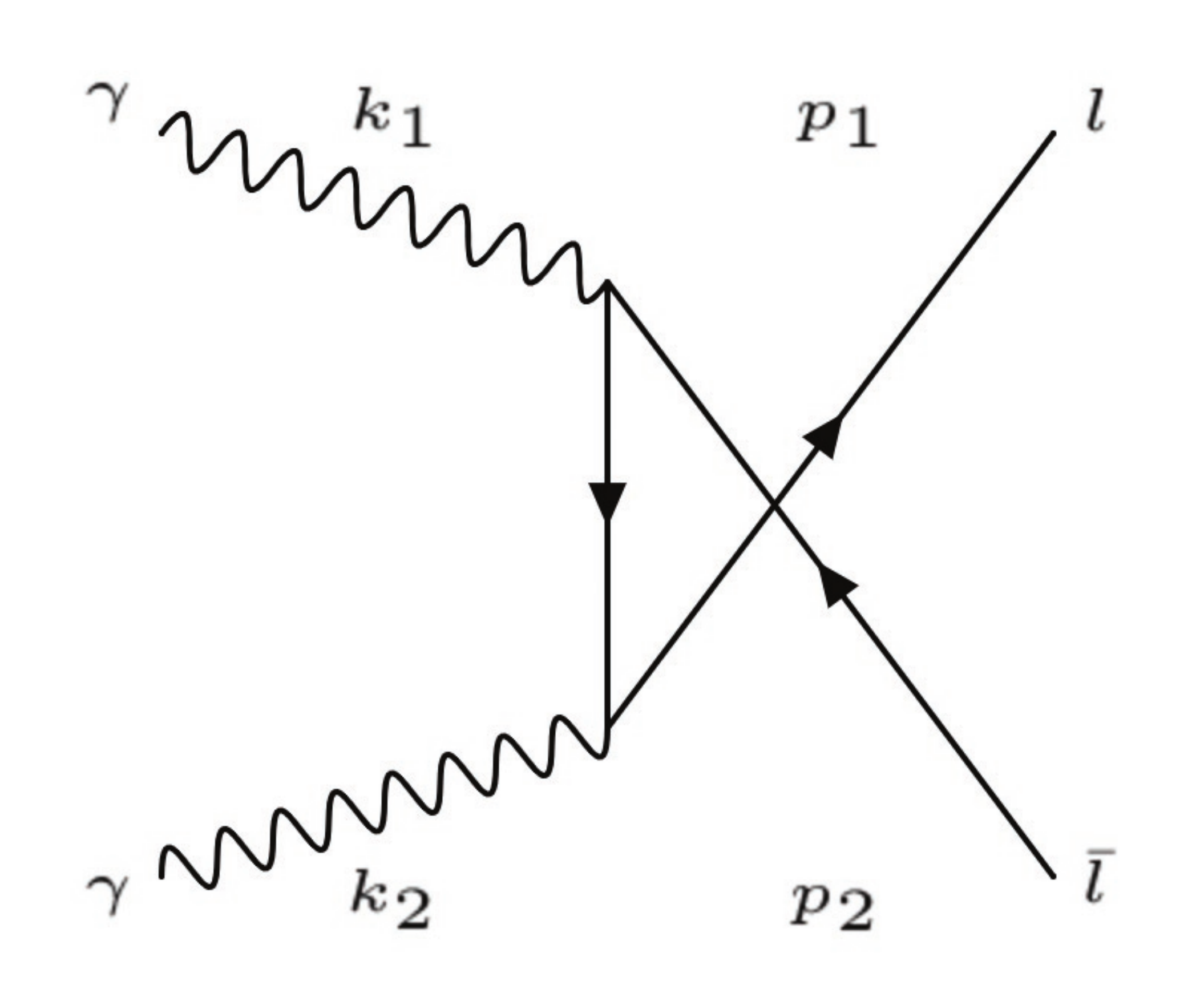}       
\end{center}
\caption{\label{feyLabel} 
 Feynman diagrams in the $t$- and $u$-channel for the process $\gamma\gamma \to l\bar{l}$.}
\end{figure}

Starting from lepton production, in order to get a feeling of what happens in different energy regimes, we recalculate the cross section $\sigma$ of the process
$\gamma\gamma \to l\bar{l}$, where $l$ is a generic charged lepton of mass $m$. The process 
$\gamma\gamma \to l\bar{l}$ receives contributions from $t$- and $u$-channel
diagrams. We compute $\sigma_{\gamma\gamma \to l\bar{l}}$ in ${\cal S}_{\rm CM}$ where the two incident photons have four-momenta $k_1\equiv(\omega, \bm{\omega})$ and $k_2\equiv(\omega, - \bm{\omega})$ and the outgoing leptons have four-momenta $p_1\equiv(\omega, {\bf p})$ and $p_2\equiv(\omega,- {\bf p})$, with 
$\omega$ denoting the photon energy and ${\bf p}$ the lepton three-momentum (see Fig.~\ref{feyLabel}).
Integrating over the total solid angle we get the total cross section~\cite{brodsky-etal}
\begin{eqnarray}
\sigma_{\gamma\gamma \to l\bar{l}}(\omega,p)=
  \frac{\pi\alpha^2}{\omega^2}\Bigg\{ \left( 1+\frac{m^2}{\omega^2}
  -\frac{1}{2}\frac{m^4}{\omega^4}\right) \, {\rm ln}\left[ \frac{(\omega+p)^2}{m^2} \right]
  -\frac{p}{\omega}\left(1+\frac{m^2}{\omega^2} \right) \Bigg\}~,  \label{sigma} 
\end{eqnarray}
where $\alpha$ is the fine-structure constant.

\begin{figure}     
\begin{center}
\includegraphics[width=.9\textwidth]{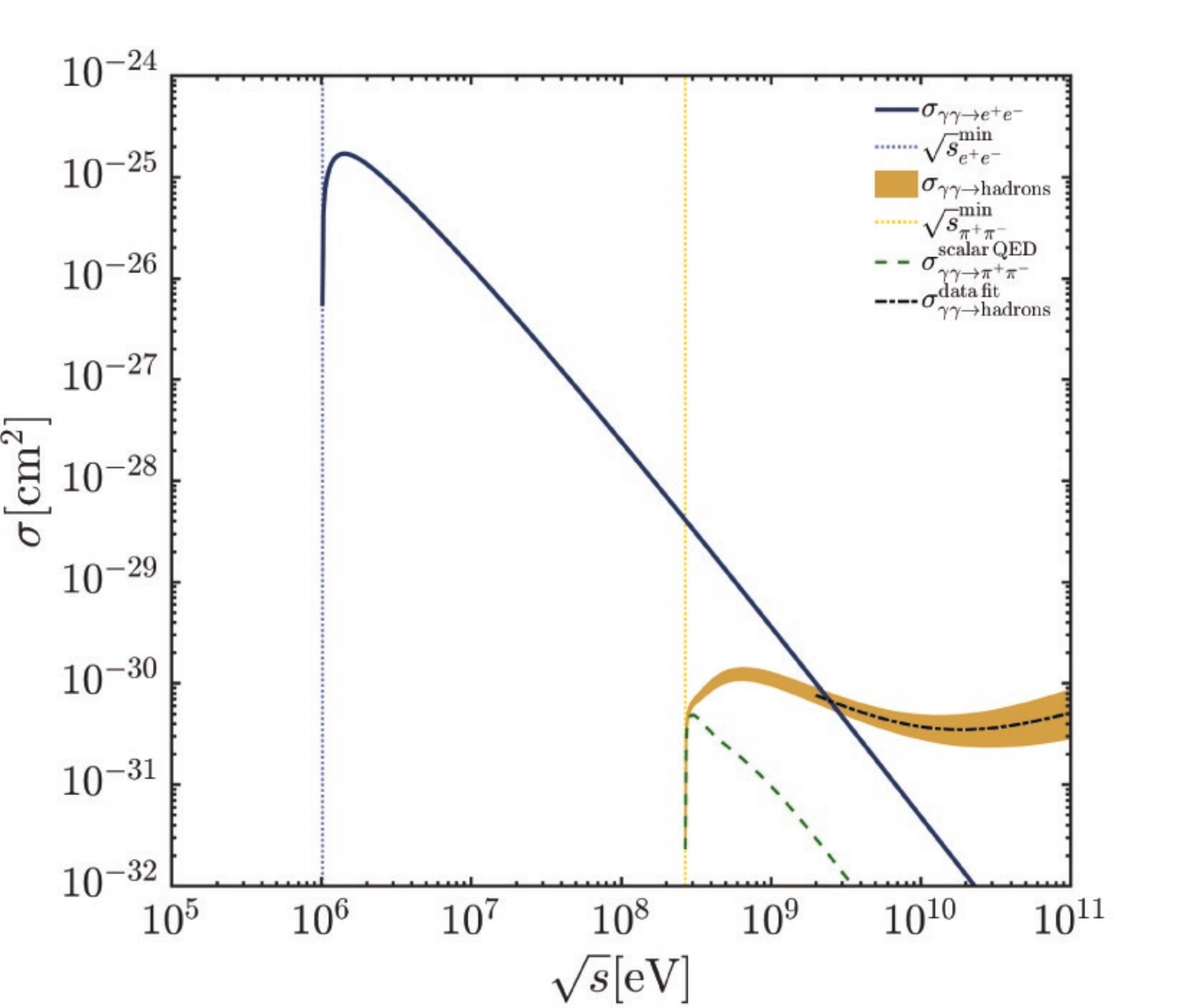}
\end{center}
\caption{\label{sigmaFig} 
  Behavior of $\sigma_{\gamma\gamma \to e^+e^-}$ and $\sigma_{\gamma\gamma \to \rm hadrons}$
  cross sections. Reported are also the thresholds above which the processes are allowed.
  The threshold for the process $\gamma\gamma \to \rm hadrons$ is such that a $\pi^+ \pi^-$ pair can be  produced, since pions are the lowest massive producible hadrons. The colored area
  represents the uncertainty affecting $\sigma_{\gamma\gamma \to \rm hadrons}$ at low energies
  and that due to data from colliders for $\sqrt{s} \gtrsim 1.5 \, \rm GeV$ in the CM frame.
  Also plotted are $\sigma_{\gamma\gamma \to \pi^+\pi^-}^{\rm scalar \, QED}$ and the data
  fit cross section $\sigma_{\gamma\gamma \to \rm hadrons}^{\rm data \, fit}$~\cite{godbole03}.
  See the text for details.}
\end{figure}

The lepton dispersion relation $\omega^2=p^2+m^2$ allows us to express 
$\sigma_{\gamma\gamma \to l\bar{l}}$ of Eq.~(\ref{sigma}) in terms of $\omega$ only, and employing the Lorentz-invariant Mandelstam variable $s=(k_1+k_2)^2=4\,\omega^2$ we can rewrite $\sigma_{\gamma\gamma \to l\bar{l}}$ in terms of $s$ alone. For instance, the behavior of $\sigma_{\gamma\gamma \to l\bar{l}}$ for the $e^+e^-$ production is reported in Fig.~\ref{sigmaFig}, while its analytic expression at high energies reads
\begin{equation}
\label{sigmaHE}
\sigma_{\gamma\gamma \to l\bar{l}}(s){\Big{|}}_{s \gg m^2} \simeq
\frac{4\pi\alpha^2}{s}\left[ {\rm  ln}\left( \frac{s}{m^2} \right)-1\right]~.
\end{equation}

As far as the production of hadrons is concerned, the picture is much more involved because of the presence of non-perturbative effects. First detailed studies of the process $\gamma \gamma \to $~hadrons in the context of $e^+ e^-$ collider physics have been presented in~\cite{brodsky-etal,budnev-etal}, while first investigations on the $\gamma \gamma$  total hadronic cross section and its relevance for the absorption of extragalactic 
$\gamma$-rays has been briefly discussed in~\cite{terasaki}. In the low-energy region the hadronic cross section is dominated by pion pair production. In addition to tree-level results based on scalar QED for charged pions, 
refined predictions in chiral perturbation theory have been computed at one- and two-loop level  in~\cite{bijnens88}  and \cite{burgi,gis1}, respectively. These predictions are reliable in the region $\sqrt{s} \leq 500$~MeV with the higher order corrections increasing the cross section by about 20\%. Also the process 
$\gamma \gamma \to \pi^0 \pi^0$ is known at one- and two-loop level,~\cite{bijnens88,dhl} and \cite{bgs,gis2}, 
respectively, with the total contribution not exceeding the cross section for charged pion production by about 
10\%. The peak of the total cross section (charged plus neutral pion pairs) shows up at $\sqrt{s} \simeq 300$~MeV, with a value of about 600~nb.

The energy region $500$~MeV $\leq \sqrt{s} \gtrsim 1.5$~GeV is populated by hadronic resonances with 
$J^{PC} = 0^{++}$~\cite{PDG}, which couple to the initial photon pair and complicate the picture. The cross section has been measured at flavor factories for different exclusive channels, with event selection cuts on the polar angle ofthe hadronic decay products (see e.g. \cite{belle1,belle2,belle3}).

The total hadronic $\gamma \gamma$ cross section in the energy region $\sqrt{s} \gtrsim 1.5$~GeV can be described by various models with their inherent uncertainties (for a review see~\cite{godbole03} and references therein). However, for CM energies up to $200$~GeV -- thus spanning two orders of magnitude in energy --  
the $\gamma \gamma$ hadronic cross section has been measured at LEP~\cite{l3,opal} and a Regge inspired parametrization of the type $\sigma_{\gamma \gamma \to \rm hadrons}^{\rm data \, fit}(s) =
A s^\epsilon + B s^{-\eta}$ has been fitted to the data. 

Our predictions for the total hadronic $\gamma \gamma$ cross section in the region $\sqrt{s} \geq 1.5$~GeV are obtained from the latter parametrization, with the following parameter values: $A=51 \pm 14 \, {\rm nb}$, $B=1132 \pm 158 \, {\rm nb}$, $\epsilon=0.240 \pm 0.032$ and $\eta=0.358$. In the low-energy regime we stick to the tree-level scalar QED prediction~\cite{brodsky-etal}
\begin{eqnarray}
\sigma_{\gamma\gamma \to \pi^+\pi^-}^{\rm scalar \, QED}(\omega,p)=\frac{\pi\alpha^2}{4\omega^2}\Bigg\{ 2 \left(1 + \frac{m_{\pi^{\pm}}^2}{\omega^2}  \right)\frac{p}{\omega} -\frac{m_{\pi^{\pm}}^2}{\omega^2}\left(2-\frac{m_{\pi^{\pm}}^2}{\omega^2} \right) {\rm ln}\left[ \frac{(\omega+p)^2}{m_{\pi^{\pm}}^2} \right] \Bigg\}~, \label{sigmaScalQED} 
\end{eqnarray}
where $m_{\pi^{\pm}}$ is the $\pi^{\pm}$ mass. We multiply the cross section by a K-factor in order to reproduce the prediction of two-loop chiral perturbation theory of about $600$~nb at $\sqrt{s}\simeq 300$~MeV. We then smoothly connect these predictions in the intermediate region, including the shape due to the hadronic resonances within our uncertainty band. The resulting cross section as a function of the CM energy is plotted in Fig.~\ref{sigmaFig}, with the threshold given by $2 \, m_\pi$. 

In order to evaluate the optical depth $\tau_{\gamma}$ of UHE photons interacting with soft background  photons, we work in a generic inertial reference frame ${\cal S}$, wherein $E$ denotes the energy of the hard photon and $\epsilon$ that of the soft background photon. As a result, $s$ reads
\begin{equation}
\label{generalRef}
s=4 \, \omega^2=2 E \epsilon(1-{\rm cos}\,\varphi)~,
\end{equation}
where $\varphi$ is the angle between the two photon three-momenta. In order to translate $\sigma_{\gamma\gamma \to l\bar{l}}$ and $\sigma_{\gamma\gamma \to \rm hadrons}$ from ${\cal S}_{\rm CM}$ to ${\cal S}$, we use the fermion dispersion relation and Eq.~(\ref{generalRef}), thereby obtaining them as written in ${\cal S}$ and denoted by $\sigma_{\gamma\gamma \to l\bar{l}}(E,\epsilon,\varphi)$ and $\sigma_{\gamma\gamma \to \rm hadrons}(E,\epsilon,\varphi)$, respectively, which will be used to compute the corresponding contribution $\tau_{\gamma,\gamma\gamma \to l\bar{l}}$ and $\tau_{\gamma,\gamma\gamma \to \rm hadrons}$ to $\tau_{\gamma}$.


\section{Double pair production (DPP)}
At very high energies the $e^+ e^-$ double pair-production cross section can be well
approximated by~\cite{originalDDP,kapusta} 
\begin{equation}
\label{sigmaDPP}
\sigma_{\rm DPP}(s)=\frac{\alpha^4}{\pi m_e^2}\left( \frac{175}{36}\zeta(3)-\frac{19}{18} \right) \left( 1- \frac{16 \, m_e^2}{s} \right)^6~,
\end{equation}
where $m_e$ is the electron mass and $\zeta$ is the zeta function. The $4 \, m_e$ threshold has been taken into account by means of a step function. This simplified approach becomes unreliable in the region close to the threshold. However, we have checked that by changing the energy threshold of the step function from $s_0 = 4 \, m_e$ to $100 \, s_0$ $\tau_\gamma$ is unaffected in the energy region of the UHE photons above $10^{18}$~eV, in the frame ${\cal S}$ we are interested in. 

The asymptotic cross section for $\gamma \gamma \to e^+ e^- \mu^+ \mu^-$ has been computed  in~\cite{kss1,kss2,masujima,kapusta}, finding values of about three orders of magnitude lower than 
$\sigma_{\rm DPP} \equiv \sigma_{\gamma \gamma \to e^+ e^- e^+ e^-}$, thus completely negligible for our purposes. 


\section{Single neutral meson production}
We consider here the production of the neutral mesons $\pi^0$, $\eta$, $\eta'$ and $\eta_c$ induced by the $\gamma\gamma$ scattering.

For $\pi^0$ the dominant decay mode is into two photons $\pi^0 \to \gamma\gamma$. Here, we consider the inverse process of single $\pi^0$ production $\gamma\gamma \to \pi^0$ whose cross section is
\begin{equation}
\label{sigmaPi0}
\sigma_{\gamma\gamma \to \pi^0}(s)=\frac{8\pi^2}{m_{\pi^0}}\Gamma_{\pi^0 \to \gamma\gamma}\delta\left( s-m_{\pi^0}^2 \right)~,
\end{equation}
where $\Gamma_{\pi^0 \to \gamma\gamma}=7.82 \, \rm eV$ is the experimental $\pi^0$ decay rate~\cite{PDG}. Obviously, the $\pi^0$ production cross section includes a Dirac delta function because of four-momentum conservation: in ${\cal S}_{\rm CM}$ the total energy of the incident photons must be exactly equal to $m_{\pi^0}$.

For $\eta$, $\eta'$ and $\eta_c$ the situation is similar, apart from the fact that not always the $\gamma\gamma$ channel represents the dominant decay mode. In Eq.~(\ref{sigmaPi0}) we must replace $m_{\pi^0}$ and $\Gamma_{\pi^0 \to \gamma\gamma}$ with the corresponding quantities for $\eta$, $\eta'$ and $\eta_c$. For the masses we take $m_{\eta}=548 \, \rm MeV$, $m_{\eta'}=958 \, \rm MeV$ and $m_{\eta_c}=2984 \, \rm MeV$ 
while for the $\gamma\gamma$ decay rates $\Gamma_{\eta \to \gamma\gamma}=0.51 \, \rm keV$, $\Gamma_{\eta' \to \gamma\gamma}=4.3 \, \rm keV$ and $\Gamma_{\eta_c \to \gamma\gamma}=5 \, \rm keV$~\cite{PDG}.

The cross section for the production of {\it para}-positronium ($p$-Ps) via $\gamma \gamma \to p{\textit -}{\rm Ps}$ possesses the same functional form of Eq.~(\ref{sigmaPi0}) where we replace $m_{\pi^0}$ with $m_{p{\textit -}{\rm Ps}}\simeq 2 \, m_e$ and $\Gamma_{\pi^0 \to \gamma\gamma}$ with $\Gamma_{p{\textit -}{\rm Ps} \to \gamma\gamma}=5.29 \times 10^{-6} \, \rm eV$~\cite{positr}.
 

\section{Optical depth} 
The optical depth ${\tau}_{\gamma}(E_0,z_s)$ at redshift $z_s$ for a hard photon of energy $E=(1+z_s)E_0$ -- $E_0$ being the observed energy in ${\cal S}$ -- interacting with background soft photons of energy $\epsilon$ is computed by multiplying the background spectral number density $n_{\gamma}({\epsilon}(z), z)$ by the cross section of two interacting photons ${\sigma}_{\gamma \gamma} (E(z), {\epsilon}(z), \varphi)$, and next integrating over $z$, $\varphi$ and ${\epsilon}(z)$~\cite{tau1,tau2}. The result is
\begin{eqnarray}
\label{eq:tau}
\tau_{\gamma}(E_0, z_s) = \int_0^{z_s} {\rm d} z ~ \frac{{\rm d} l(z)}{{\rm d} z} \, \int_{-1}^1 {\rm d}({\cos \varphi}) ~ \frac{1- \cos \varphi}{2} \ 
\times \\
\nonumber
\times  \, \int_{\epsilon_{\rm thr}(E(z) ,\varphi)}^\infty  {\rm d} \epsilon(z) \, n_{\gamma}(\epsilon(z), z) \,  
\sigma_{\gamma \gamma} \bigl( E(z), \epsilon(z), \varphi \bigr)~, \ \ 
\end{eqnarray}
where
\begin{equation}
\label{lungh}
\frac{{\rm d} l(z)}{{\rm d} z} = \frac{1}{H_0} \frac{1}{\left(1 + z \right) \left[ {\Omega}_{\Lambda} + {\Omega}_M \left(1 + z \right)^3 \right]^{1/2}}~.
\end{equation}
In the standard $\Lambda$CDM cosmological model we take for definiteness $H_0 = 70 \, {\rm km} \, {\rm s}^{-1} \, {\rm Mpc}^{-1}$ -- which in natural units reads $H_0 = 1.50 \times 10^{ - 33} \, {\rm eV}$ -- while ${\Omega}_{\Lambda} = 0.7$ and ${\Omega}_M = 0.3$. In addition, we have
\begin{equation} 
\label{eq.sez.urto01012011}
{\epsilon}_{\rm thr}(E,\varphi) \equiv \frac{m_{\rm thr}^2}{2 E  \left(1-\cos \varphi \right)}~,
\end{equation}
where $m_{\rm thr}$ is the total mass of the produced particles: for leptons $m_{\rm thr}$ is the mass of two/four leptons, for hadrons $m_{\rm thr}$ is the mass of the two produced mesons, the minimum being $2 \, m_\pi$~\cite{thr}.

Once $n_{\gamma}(\epsilon(z), z) $ is known, $\tau_{\gamma}(E_0, z)$ can be computed exactly, but generally the integration over $z$ in Eq. (\ref{eq:tau}) must be performed numerically. Concerning $n_{\gamma}(\epsilon(z), z)$ of the soft photon background we consider the EBL, the CMB and the RB. Many methods exist in the literature to estimate the EBL background but they substantially agree in the redshift range where they overlap, so that nowadays the EBL is known to a very good accuracy. We adopt here the model by Gilmore et al.~\cite{EBLdom} mainly because the values of $n_{\gamma}(\epsilon(z), z)$ are tabulated. Similar results can be obtained e.g. by employing the model of Franceschini \& Rodighiero~\cite{EBLfra}. For the CMB we consider the standard temperature value $T = 2.73 \, \rm K$, and concerning the RB we use the most recent available data with a low-frequency cutoff placed at $2 \, \rm MHz$~\cite{RBgerv}.


\section{Results}
 We are now in position to evaluate the contribution to the optical depth $\tau_{\gamma}$ arising from all considered processes, namely $\tau_{\gamma,\gamma\gamma \to l\bar{l}}$ ($l=e,\mu,\tau$), 
$\tau_{\gamma,\gamma\gamma \to {\rm hadrons}}$, $\tau_{\gamma,\gamma\gamma \to e^+e^-e^+e^-}$, $\tau_{\gamma,\gamma\gamma \to \pi^0,\eta,\eta',\eta_c}$ and $\tau_{\gamma,\gamma\gamma \to p{\textit -}{\rm Ps}}$.

\begin{figure}      
\begin{center}
\includegraphics[width=.9\textwidth]{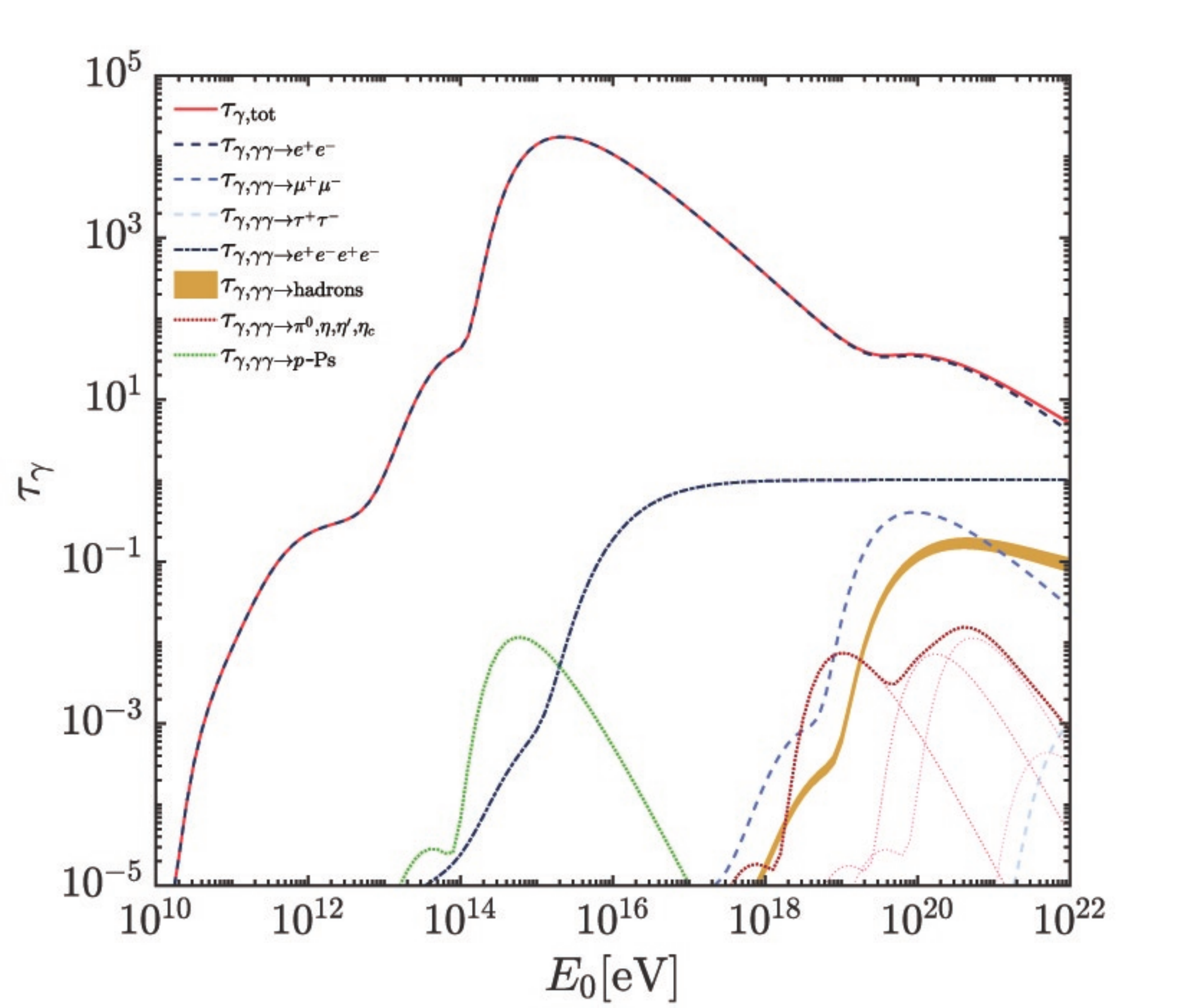}
\end{center}
\caption{\label{tauTot} 
Optical depth for all considered processes for a source at $z_s=0.03$ ($\simeq 130 \, \rm Mpc$). The colored area represents the uncertainty on the optical depth corresponding to the $\gamma\gamma\to \rm hadrons$ channel caused by the uncertainty affecting $\sigma_{\gamma \gamma \to \rm hadrons}$. See the text for more details.}
\end{figure}

In Fig.~\ref{tauTot} we plot the optical depth at $z_s=0.03$ (corresponding to a distance $\simeq 130 $ Mpc) of all processes considered so far, along with the total optical depth $\tau_{\gamma,{\rm tot}}$. The shadowed area represents the uncertainty of $\tau_{\gamma}$ arising from that concerning $\sigma_{\gamma \gamma \to \rm hadrons}$. 

At energies $E_0 \gtrsim 10^{18} \, \rm eV$ $\tau_{\gamma,{\rm tot}}$ starts to differ {\it in a sizable way} from the optical depth arising from the process $\gamma\gamma \to e^+e^-$ alone, which represents the dominant contribution at almost all energies. As we said, in the literature also $\gamma\gamma \to e^+e^-e^+e^-$ has been considered (see e.g. ~\cite{TauDPP}): this process starts to become important at energies $E_0 \gtrsim 10^{18} \rm eV$. 

\begin{figure}      
\begin{center}
\includegraphics[width=.9\textwidth]{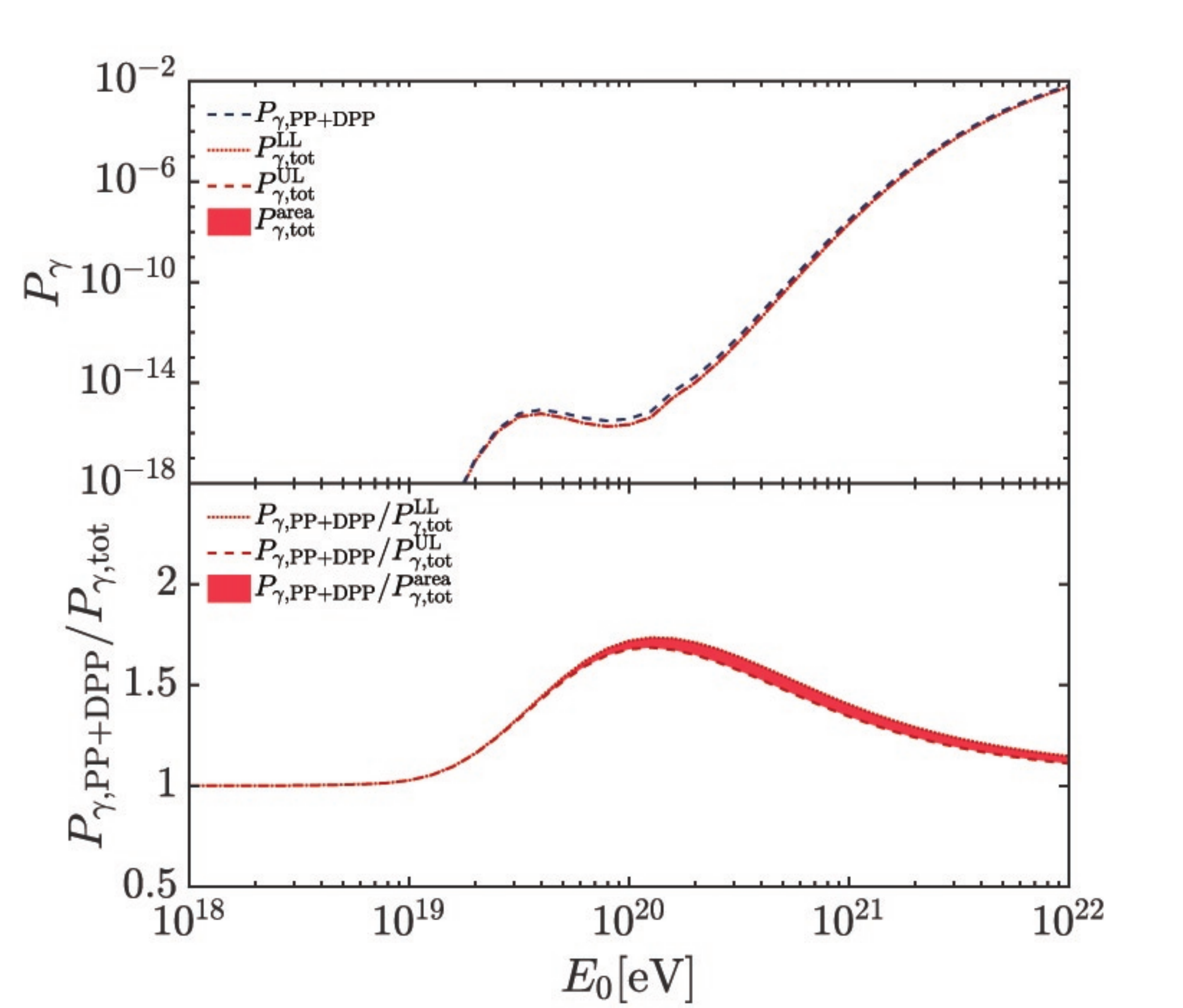}
\end{center}
\caption{\label{ratioProb}Survival probability $P_{\gamma}$ for UHE photons at $z_s=0.03$ ($\simeq 130 \, \rm Mpc$). In the upper panel the total photon survival probability $P_{\gamma, {\rm tot}}$ (with very small area of uncertainty caused by uncertainty affecting $\sigma_{\gamma \gamma \to \rm hadrons}$ and lower and upper limits $P_{\gamma, {\rm tot}}^{\rm LL}$ and $P_{\gamma, {\rm tot}}^{\rm UL}$) and the photon survival probability $P_{\gamma, {\rm PP+DPP}}$ -- when only $\gamma\gamma \to e^+e^-$ and $\gamma\gamma \to e^+e^-e^+e^-$ are considered -- are plotted. In the lower panel the ratio between $P_{\gamma, {\rm PP+DPP}}$ and $P_{\gamma, {\rm tot}}$ is drawn with the same conventions of the upper panel.}
\end{figure}

In order to figure out the importance of the considered processes besides $\gamma\gamma \to e^+e^-$ and $\gamma\gamma \to e^+e^-e^+e^-$, it is compelling to compute the photon survival probability $P_{\gamma}$ -- which is related to $\tau_{\gamma}$ by $P_{\gamma}=e^{-\tau_{\gamma}}$ -- since this is the actually observable quantity. In the upper panel of Fig.~\ref{ratioProb} we plot the total photon survival probability $P_{\gamma, {\rm tot}}$  taking into account all considered processes, and the photon survival probability $P_{\gamma, {\rm PP+DPP}}$ when only $\gamma\gamma \to e^+e^-$ and $\gamma\gamma \to e^+e^-e^+e^-$ are considered. For $P_{\gamma, {\rm tot}}$ we plot also the shadowed area with a similar meaning of Fig.~\ref{tauTot}. The lower and upper limits $P_{\gamma, {\rm tot}}^{\rm LL}$ and $P_{\gamma, {\rm tot}}^{\rm UL}$ are also shown: they are associated with the highest and lowest value of $\sigma_{\gamma \gamma \to \rm hadrons}$, respectively. The lower panel of Fig.~\ref{ratioProb} uses the same conventions of the upper one and quantifies  the importance of the new considered processes with respect to $\gamma\gamma \to e^+e^-$ and $\gamma\gamma \to e^+e^-e^+e^-$ alone. Thus, in this panel we plot the ratio between $P_{\gamma, {\rm PP+DPP}}$ and $P_{\gamma, {\rm tot}}$. The reader might well be confused at this point, since $P_{\gamma, {\rm tot}}$ is smaller than $P_{\gamma, {\rm PP+DPP}}$, which looks like a paradox. Let us explain why it is 
not. We recall that $P_{\gamma, {\rm tot}}$ has been defined as the photon survival probability, namely the probability that the incoming photons do not undergo any process considered above. Yet, $P_{\gamma, {\rm PP+DPP}}$ is the probability that such photons do not convert into $e^+ e^-$ or $e^+e^-e^+e^-$. So, it is the quantity 
$1 - P_{\gamma, {\rm tot}}$ that cannot be smaller than $1 - P_{\gamma, {\rm PP+DPP}}$.

We see that the contribution of the processes considered in this Letter is not negligible in the UHE band since at redshift $z_s=0.03$ ($\simeq 130 \, \rm Mpc$) it gives a correction of about $2\%$ to the total optical depth $\tau_{\gamma,{\rm tot}}$, which translates into a decrease of the total survival probability $P_{\gamma, {\rm tot}}$ by a factor of about 2.

In particular, as we can see from Fig.~\ref{tauTot} the most important channels in addition to $\gamma\gamma \to e^+e^-$ and $\gamma\gamma \to e^+e^-e^+e^-$ are those leading to the production of hadrons and of $\mu^{\pm}$ , while the production of $\tau^{\pm}$ is less important. The single neutral meson production that takes into account $\pi^0$, $\eta$, $\eta'$ and $\eta_c$ does not play any significant role as compared to the above-mentioned processes: only around $E_0\sim 5 \times10^{18} \, \rm eV$ represents it the leading contribution among those that must be added to the processes $\gamma\gamma \to e^+e^-$ and $\gamma\gamma \to e^+e^-e^+e^-$. The {\it para}-positronium production is the most important channel after 
$\gamma\gamma \to e^+e^-$ around $10^{14} \, {\rm eV} \lesssim  E_0\lesssim 10^{15} \, {\rm eV}$ but its correction to the process $\gamma\gamma \to e^+e^-$ is totally irrelevant for UHE photons.


\section{Discussion and conclusion}
In this Letter we have evaluated the relevance for the transparency of UHE photons of hadron- and lepton-photoproduction reactions, and the single neutral meson and {\it para}-positronium production process, on top of the $\gamma\gamma \to e^+e^-$ and $\gamma\gamma \to e^+e^-e^+e^-$ processes. As far as the hadron sector is concerned, uncertainties affecting $\tau_{\gamma,\gamma\gamma \to \rm hadrons}$ give rise to an uncertainty in $\tau_{\gamma,{\rm tot}}$ or equivalently in $P_{\gamma, {\rm tot}}$. Yet, our results are sizable even in the most conservative case (taking the smallest value of $\sigma_{\gamma\gamma \to \rm hadrons}$). Specifically, $\tau_{\gamma,{\rm tot}}$ changes by about $2\%$ and $P_{\gamma, {\rm tot}}$ is reduced by a factor of about $2$ at $z_s=0.03$.

A remark is now in order. It goes without saying that any final product of the two-photon scattering triggers an electromagnetic cascade, producing secondary photons and other particles (see e.g.~\cite{EMcascade}). However, the study of this topic is beyond the scope of this Letter, since here our attention is focussed on the UHE photon survival probability.

In conclusion, we restate that our calculation of the cosmic photon opacity -- by including the most important interaction channels for the process $\gamma\gamma \to {\rm any}$ -- is extremely important for the evaluation of upper limits on the UHE photon flux (see e.g.~\cite{auger}), and the existing ones should be modified according to our findings.

\ \

\begin{acknowledgments}

We thank Fr\'ed\'eric Kapusta for useful information. G. G. and F. T. acknowledge contribution from the grant INAF CTA--SKA, ``Probing particle acceleration and $\gamma$-ray propagation with CTA and its precursors" and the INAF Main Stream project ``High-energy extragalactic astrophysics: toward the Cherenkov Telescope Array''. The work of F. P. is supported by an INFN QFT@Colliders grant. The work of M. R. is supported by an INFN TAsP grant. 
\end{acknowledgments}

\end{document}